\documentclass[runningheads]{llncs}
\usepackage{silence}
\usepackage[T1]{fontenc}
\usepackage{algorithm}
\usepackage[noend]{algpseudocode}
\usepackage{graphicx}
\usepackage{subcaption}
\usepackage{delarray}
\usepackage{xurl}

\usepackage{comment}
\usepackage{epstopdf}

\usepackage{amsmath}

\begin{document}

\title{Improving Capstone Team Outcomes through Dynamic Skill Matching and Preference Alignment}
\titlerunning{Improving Capstone Team Formation}

\author{
Brandon Pardi \inst{1} \and
Garret Castro \inst{2} \and
Michael Pisman \inst{2} \and
Avash Adhikari \inst{2} \and
Santosh Chandrasekhar\inst{2}
}

\authorrunning{Pardi et al.}

\institute{University of California, Davis, Davis, CA 95616, USA\and University of California, Merced, Merced, CA 95343, USA}

\maketitle              

\begin{abstract}

Team-based projects are a cornerstone of engineering and computing courses, but unstructured team formation often leads to poor project outcomes due to misaligned student interests and inadequate skill coverage. This paper introduces a novel, three-stage methodology for creating effective student teams by integrating student preferences with project skill requirements.

In the first stage, students complete a survey to report their project interests and self-assessed skills. Next, a Large Language Model (LLM) analyzes project descriptions to extract the necessary skills for each project’s success. Finally, a dynamic assignment algorithm matches students to projects, simultaneously maximizing skill coverage and preference alignment. The algorithm iteratively prioritizes projects with unfulfilled skill needs to optimize team balance.

Preliminary evaluations show our approach produces teams with higher skill coverage and better preference satisfaction compared to
random or manual assignment approaches. Our approach also overcomes limitations of widely-used tools like CATME Team-Maker, which do not explicitly account for project skill fulfillment. Our findings point toward an effective and customizable strategy for improving student motivation and learning outcomes in project-based courses.


\end{abstract}

\section{Introduction}
\label{sec:intro}

Capstone and senior design courses in engineering and computing education are typically organized around project-based learning, providing students with opportunities to collaboratively apply their acquired knowledge and skills to authentic and professionally relevant problems. These courses often involve teams working on real-world projects that are proposed by industry or academic sponsors. The success of such project-based courses strongly relies on effective team formation to ensure that student teams possess both the necessary skills to successfully complete projects and a strong interest alignment with their assigned projects to maintain student motivation and agency. However, without a carefully designed assignment process, students often end up assigned to projects that either do not match their interests or lack critical skill coverage, which can result in reduced motivation, lack of agency, and poor project outcomes~\cite{kessler2025hierarchical}.

In practice, team formation methods can be broadly classified into student-led and instructor-led approaches. While student-led team formation can lead to increased agency, it often produces suboptimal results in scenarios where projects have predefined skill requirements and objectives~\cite{oakley2004turning}. More specifically, student-led approaches typically fail to ensure that students are matched with projects that are aligned with their skill sets, thus negatively impacting project success. On the other hand, instructor-led team formation methods can be more effective in aligning student skills and preferences with explicit project requirements. However, when done manually, an instructor-led process is both labor-intensive and susceptible to bias, particularly when involving complex and nuanced metrics such as detailed skill sets and individual student preferences~\cite{parker2019launching}. These challenges have led to a growing interest in data-driven or automated methods to streamline and improve team formation.

Existing automated tools, such as CATME Team-Maker~\cite{loughryDesignValidationWebBased2010}, address several common challenges of team formation by grouping students based on demographic factors and personality traits. However, CATME lacks explicit mechanisms for detailed skill-to-project matching and does not dynamically integrate student preferences into the assignment process. This necessitates manual intervention post-assignment, adding significant cognitive load on instructors and potentially reducing the effectiveness of teams~\cite{hastingsComposingTeamCompositions2023}. Other methods employing optimization techniques, heuristic algorithms, or machine learning approaches, while capable of forming balanced teams, similarly fail to simultaneously address detailed project-specific skill coverage and student preferences dynamically. Furthermore, approaches relying purely on clustering methods or stable matching algorithms typically focus only on optimizing student satisfaction without adequately considering the detailed skill needs of each project. Recent advances in natural language processing, particularly the emergence of large language models (LLMs), offer promising solutions for automating skill extraction from project descriptions to facilitate more accurate team formation.

In this paper, we propose a novel, data-driven team formation algorithm that dynamically balances detailed project-specific skill requirements with individual student preferences. Our approach comprises of three stages: (1) student surveys to capture detailed and granular skill sets and project preferences; (2) LLM-driven extraction of essential project skills from sponsor-provided descriptions; and (3) a dynamic matching algorithm that iteratively prioritizes currently unfulfilled skills and integrates student preferences during team formation. To systematically evaluate our approach, this study addresses the following research questions:

\begin{description}
    \item[RQ1:] Does dynamic skill weighting increase project skill coverage compared to static weighting approaches?
    \item[RQ2:] Does our algorithm improve student preference satisfaction compared to manual and random assignment baselines?
    \item[RQ3:] Does the combined approach of skill-preference matching outperform existing team formation methods in terms of overall team formation quality?
\end{description}

We validate the effectiveness of our approach using real data from capstone courses at University of California, Merced (UC Merced), collected during the Fall 2023 and Spring 2024 semesters. Analysis of our results indicates that our algorithm significantly improves skill coverage compared to manual and random assignments, while maintaining strong alignment with student preferences, demonstrating its potential to enhance student motivation and agency, and improve project outcomes in capstone courses.

\smallskip\noindent{\bf Contributions.} To summarize, the main contributions of our work are as follows:
\begin{itemize}
    \item A novel three-stage methodology for automated team formation that combines student skill surveys, LLM-based project skill extraction, and dynamic preference-aware matching.
    \item An LLM-driven approach for automatically extracting essential technical skills from project descriptions, enabling more accurate skill-to-project alignment.
    \item A dynamic matching algorithm that iteratively prioritizes unfulfilled project skill requirements while simultaneously optimizing student preference satisfaction.
    \item Empirical validation on real capstone course data demonstrating significant improvements in skill coverage and preference alignment compared to manual and random assignment methods.
\end{itemize}

\section{Related Work}
\label{sec:related}

The problem of forming effective student teams in capstone and senior project courses has been widely studied, leading to a variety of algorithmic approaches ranging from rule-based heuristics to optimization and machine learning techniques. While existing tools provide structured methods for team assignment, they often focus on student attributes within teams. Moreover, these tools lack the ability to simultaneously perform fine-grained skill matching and dynamically integrate student preferences, both of which are crucial for successful team formation in capstone settings.

Recent work has further highlighted that many published team-formation algorithms narrowly target specific formulations (e.g., a fixed objective mix, a single preference model, or rigid constraints), limiting generalizability across diverse courses. This fragmentation makes it difficult for instructors to compare methods or adapt them to evolving pedagogical goals \cite{kessler2025hierarchical}.

Additionally, recent advances in natural language processing (LLMs) enable automated extraction of project-required skills from textual descriptions~\cite{decorteExtremeMultiLabelSkill2023,deusserInformedNamedEntity2023,kwakBridgingLargeLanguage2024}. This section reviews prior work on tools and algorithms, the role of LLMs, and key research gaps our approach addresses.

\smallskip\noindent\textsc{CATME Team-Maker and Related Methods:}
CATME Team-Maker \cite{loughryDesignValidationWebBased2010} provides structured surveys and semi-automated grouping to balance traits (e.g., leadership, availability). However, it does not explicitly match student skills to project requirements, nor does it dynamically weight scarce skills as assignments progress, often requiring post-processing by instructors \cite{hastingsComposingTeamCompositions2023}.

In a multi-instructor study, most CATME users reported performing manual “sanity checks” and re-running configurations, citing concerns about opaque scoring, criterion weighting, default biases, and limited transparency when using sensitive attributes \cite{hastingsComposingTeamCompositions2023}. Meulbroek et al. \cite{meulbroekFormingMoreEffective2019} propose integrating Gale--Shapley \cite{gale1962college} to bring student preferences into capstone assignment, but the implementation optimizes student satisfaction only and does not provide skill-to-project matching, shifting burden to instructors. In contrast, we target both preference alignment and explicit project skill coverage.

\smallskip\noindent\textsc{Other Criterion-Based Heuristic Matching Methods:}
FASTT \cite{bulmerFASTTTeamFormation2020} uses a multi-round selection mechanism and fairness criteria to balance teams, but does not account for project preferences. Genetic-algorithm families \cite{agarwalTeamFormationEngineering2022,bergeyTeamMachineDecision2014} optimize homogeneity/diversity proxies across teams; while useful for balance, they typically do not integrate granular project skill coverage with student project choices.

\smallskip\noindent\textsc{Machine Learning Methods:}
Clustering approaches (e.g., k-means over pro- ject choices) can structure cohorts \cite{akbarImprovingFormationStudent2018} but do not guarantee per-project skill coverage. Cognitive and diagnostic models (e.g., SDINA, Bayesian roles) \cite{alberolaArtificialIntelligenceTool2016,liuCollaborativeLearningTeam2016} estimate proficiency and roles from assessment and peer data, improving team heterogeneity and dynamics, yet they require rich longitudinal data and typically do not incorporate student project preferences explicitly. ML--assisted heuristics such as MCTS pipelines introduce predictive signals (availability, proficiency growth) \cite{chenMultitaskOrientedTeam2025}, but often rely on large activity datasets \cite{feng2019understanding,kulshrestha2021web} and are rarely validated in situated capstone deployments.

\smallskip\noindent\textsc{Optimization and Integer Linear Programming (ILP) Methods:}
Optimization based methods (e.g., ILP) can provide guarantees for specific objectives (interest similarity, balance) \cite{diasNewAlgorithmCreate2017,sadeghiEffectiveGroupFormation2016}. Closer to capstone needs, Kessler et al.~\cite{kessler2025hierarchical} propose a hierarchical ILP-based method to coordinate multiple pedagogical goals, and Hammond et al.~\cite{hammond2023} incorporate student project preferences alongside skills and instructor constraints. While expressive, ILP approaches can be slow at scale and solver-dependent. In contrast, our greedy matching algorithm runs in O($n^3$), scales with cohort size, and maintains both project-specific skill fulfillment and project-choice alignment during assignment. Moreover, our algorithm is designed to minimize assignment bias and ensure equitable distribution of student expertise and interests across all projects.

\smallskip\noindent\textsc{Complexity Considerations and Hierarchical Constraints:}
Natural formulations that jointly enforce capacity, multi-skill coverage, and preference constraints are NP-hard, requiring careful objective design and practical compromises for course-scale deployments \cite{kessler2025hierarchical}. This motivates adaptive or hybrid strategies like ours that quickly approximate quality solutions while exposing tunable levers (e.g., preference weighting) to instructors.

In summary, existing methods tend to capture broad characteristics or preferences but rarely optimize both project-specific skill fulfillment and project choice simultaneously with dynamic weighting. To the best of our knowledge, our approach is one of the first to utilize LLM-derived skills and instructor constraints to combine both ``bottom-up''  and ``top-down'' methods to form teams. Our algorithm optimizes project preference and skill coverage while maintaining scalability which positions this work among the first capstone-focused methods to use both agency and fine-grained skill fulfillment in a single, scalable pipeline at classroom scale.

\section{Methodology}
\label{sec:method}
\begin{figure}[!ht]
    \centering
    \fboxsep=2pt
    \fboxrule=0.4pt
    \fbox{\includegraphics[width=0.95\linewidth]{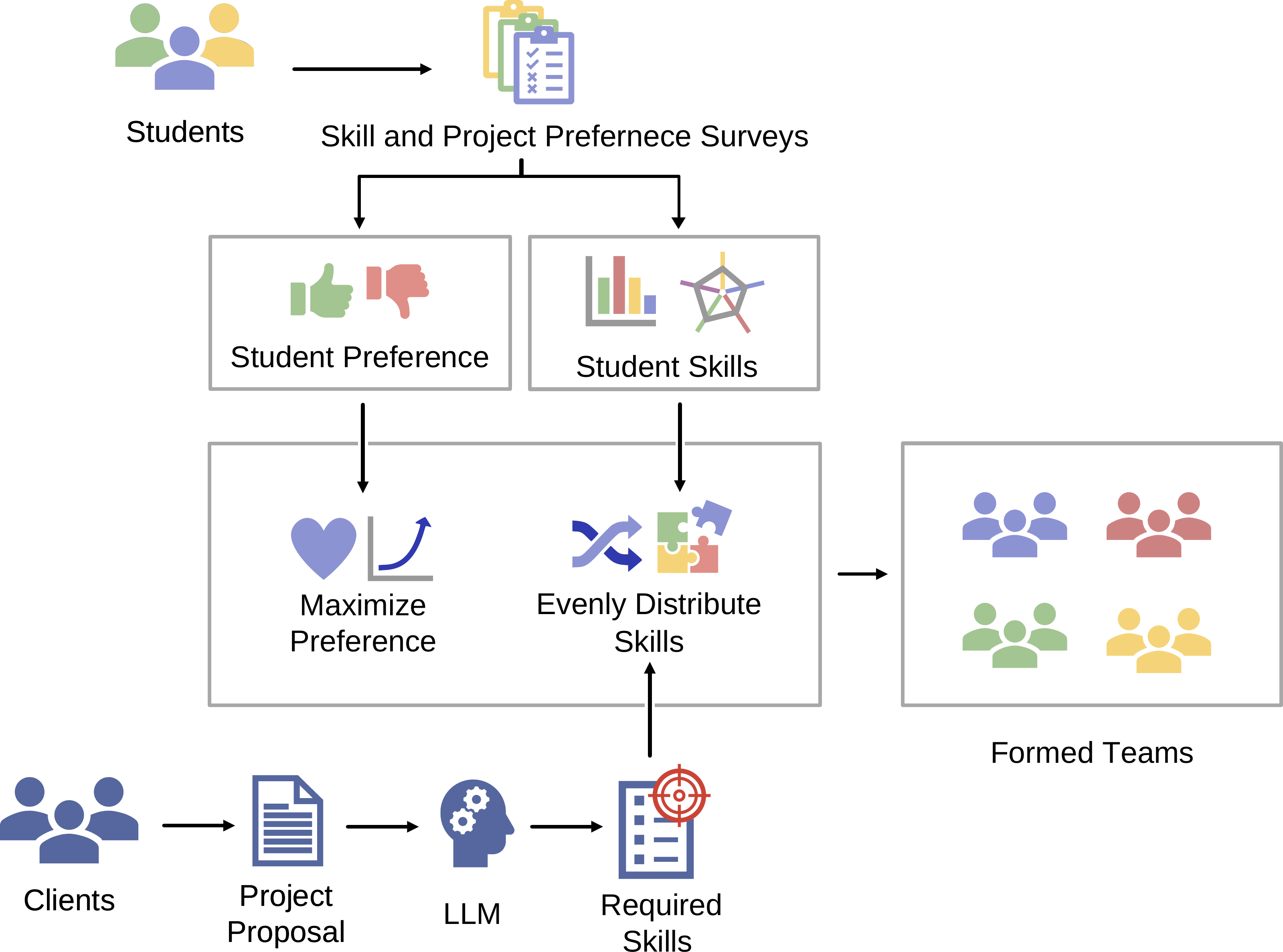}}
    \captionsetup{labelformat=empty} 
    \caption{\scriptsize Fig. 1: Flow diagram giving a generalized overview of the team assignment algorithm.}
    \label{fig:flow_diagram}
\end{figure}

Here we introduce our proposed three-step team formation process that begins with conducting a student survey to capture student skills and project preferences, followed by extraction of project skills using a LLM and finally, the matching system to form the teams while optimizing skill coverage and student preference. The general overview of our system is shown in Fig.~\ref{fig:flow_diagram}.

\subsection{Student Survey}
\label{sec:method_survey}
To effectively match students to projects, information on each student's technical skill sets and project preferences is collected through a survey at the beginning of the course. Prior to completing the surveys, all students are given access to the summary descriptions of all projects, which include the background and problem statement for each project as submitted by the sponsors. In the skill survey, students self-report and score their skills based on their experience from a predetermined list. The list of skills are selected either because projects in prior semesters had required them or to anticipate future projects needing them. Our approach can be easily adapted to work with any set of skills or student \emph{traits} as long as they can be identified as requirements for the projects for the matching system to consider in its calculations (Sec.~\ref{sec:match}). In the preference survey, students provide their preference score for all projects on a discrete scale of one to five, with five indicating the highest preference score. 

\subsection{Skill Extraction (via LLM)}
\label{sec:method_skill}

While use of LLMs for extracting specific skills in educational projects is limited, there are results that showcase their ability to find skills from text associated with educational content~\cite{kwakBridgingLargeLanguage2024}, and there are several empirical studies that show LLMs are an effective method for extracting skills from unstructured text in the professional industry~\cite{decorteExtremeMultiLabelSkill2023}. Furthermore, LLMs have also been shown to outperform previous methods such as Named Entity Recognition (NER) for extracting skills specifically from unstructured textual data~\cite{deusserInformedNamedEntity2023}. Our current implementation uses a preexisting fine-tuned instruct version of Llama 3.2~\cite{Grattafiori2024Llama3,QuantFactory_MetaLlama3_8B_Instruct_GGUF_2024} applied to a novel use case of LLM skill extraction, aiding capstone project team formation by selecting skills required for project completion. 

To identify the skills required for each project, the project descriptions are parsed and sent to the LLM along with the same skill list that is given to the students. It is then asked to rate the necessity of each skill to complete the project on a scale from 0 to 5. A 0 in this case is given when a skill doesn’t help for any plausible approach to the project, and a 5 is given when a skill is realistically indispensable for a project. Many project often have multiple possibilities of completion (e.g. one could use Pytorch or Tensorflow to achieve the same results) so the prompt includes such caveats, and instructed to consider multiple potential solutions. This workflow results in a structured JSON list of skills and their rated necessity, that is then manually reviewed and given minor or no corrections.

\subsection{Matching System}
\label{sec:match}
Our student-to-project matching algorithm works by iteratively and dynamically computing a match score for every student-project pair, reflecting the strength of their compatibility. Students are assigned to projects based on these computed scores, after which scores are recalculated to reflect updates in skill coverage and team composition. The match score integrates two \emph{weighted} parameters: student preferences and skill compatibility with project requirements. The preference weight can be adjusted by the instructor to prioritize either student preferences or skill coverage. At the same time, the skill weight is dynamically updated throughout the iterative team formation process to prioritize students who fulfill unmet skill requirements. By dynamically recalculating these weighted scores during assignment, our algorithm minimizes assignment bias in two ways. First, it prevents giving unfair advantages (in terms of project preference) to students assigned earlier. And second, it ensures that no single project disproportionately receives all the most suitable students. This way, we ensure equitable distribution of student expertise and interests across all projects.

Let $P$, $S$, and $K$ denote the sets of projects, students, and skills, respectively. Here, $K$ represents all possible skills listed in the survey described in Sec~\ref{sec:method_survey}, regardless of whether any particular project requires these skills or any student possesses them. For a student $s$ and a project $p$, we define:
\begin{description}
    \item[$PS_s(p)$: ] Student $s$'s preference score for project $p$.
    \item[$K_p \subseteq K$:] Set of all skills required for $p$.
    \item[$K_p' \subseteq K_p$:] Set of currently unfulfilled skills for $p$.
    \item[$SS_s(k)$: ] Student $s$'s self-scoring of skill $k$.
    \item[$\alpha$: ] A tunable instructor-defined parameter called the ``preference weight.'' 
    \item[$\beta_p(k)$: ] A dynamic parameter called the ``skill weight'' of skill $k$ for project $p$. 
\end{description}
The instructor sets the value of $\alpha$ to control the influence of student preferences on team assignment. For each skill $k \in K_p$, $\beta_p(k)$ is initialized to 1. Otherwise, if $|K_p'| > 0$ (there are still some unfulfilled skills for a project), $\beta_p(k)$ is computed as $|K_p|/|K_p'|$. A skill is considered \emph{unfulfilled} if none of the students currently assigned to that project have self-rated their proficiency in that skill at a level of 3 or higher (on the survey's 1--5 scale). 

Based on this notation, we define an auxiliary function to our student-to-project algorithm. The “match score” $MS_{s,p}$ between student $s$ and project $p$ is computed as follows:
\begin{equation}
\label{eqn:ms}
MS_{s,p} = \alpha \ast PS_s(p) + \sum\limits_{k \in K_p} \beta_p(k) \ast SS_s(k)
\end{equation}
Notice that as the algorithm progresses, and more students are assigned to the project, the value of $\beta_p(k)$ for unfulfilled skills will increase leading to a higher match score for students that satisfy those skills. This dynamic adjustment of the skill-weight enables better skill coverage for the projects.

\smallskip\noindent\textsc{Assignment Algorithm: } Students are iteratively assigned using a greedy process (Alg.~\ref{algo}). For each round, the highest $MS_{s,p}$ among unassigned students and available projects is selected; ties are broken by preference score, then randomly. Scores are recomputed after each assignment until all students are placed. Projects with higher average preference scores are given more capacity, up to a fixed team-size limit (five in our case). This approach ensures fairness and maximizes skill coverage.

\begin{algorithm}[hbt!]
\footnotesize
\captionsetup{labelformat=empty} 
\caption{\footnotesize \textbf{Algorithm 1} Greedy Student-to-Project Assignment with Dynamic Skill Weighting}\label{algo}
\begin{algorithmic}[1]

\Require{List of students $S$, list of projects $P$, preference weight $\alpha$}
\Ensure{Student-to-project assignments}

\While{some $s \in S$ is unassigned}
    \State $\mathsf{max\_score} \leftarrow -\infty$\, $\mathsf{best\_pair} \leftarrow$ None
    \For{unassigned student $s \in S$}
        \For{project $p \in P$ with space} 
        
            \State Let $K_p$ be required skills and $K_p'$ be the currently unfulfilled skills
            \State Initialize all $\beta_p(k) \leftarrow 1$ for $k \in K_p$
            \If{$|K_p'| > 0$} 
                \For{skill $k \in K_p'$}
                    \State $\beta_p(k) \leftarrow {|K_p|}/{|K_p'|}$
                \EndFor
            \EndIf

            \State Compute match score: $MS_{s,p} \leftarrow \alpha \ast PS_s(p) + \sum\limits_{k \in K_p} \beta_p(k) \ast SS_s(k)$

            \If{$MS_{s,p} > \mathsf{max\_score}$}
                \State $\mathsf{max\_score} \leftarrow MS_{s,p}$, $\mathsf{best\_pair} \leftarrow (s, p)$, $\mathsf{max\_pref} \leftarrow PS_s(p)$
            \State
            \ElsIf{$MS_{s,p} = \mathsf{max\_score}$} 
                \If{$PS_s(p) > \mathsf{max\_pref}$} 
                    \State $\mathsf{best\_pair} \leftarrow (s, p)$, $\mathsf{max\_pref} \leftarrow PS_s(p)$
                \ElsIf{$PS_s(p) = \mathsf{max\_pref}$}
                    \State Randomly decide whether to update $\mathsf{best\_pair}$
                \EndIf
            \EndIf
        \EndFor
    \EndFor
    \State Assign student $s$ to project $p$ from best\_pair and mark $s$ as assigned and update $p$'s assigned list
\EndWhile
\end{algorithmic}
\end{algorithm}

\noindent\textsc{On Support for Courses With Labs: } 
The capstone course at UC Merced includes both lecture and lab sessions. Outside of their self-organized meetings, student teams attend these lab sessions to meet and plan their tasks for the week. Although projects can receive students from any lab, we enforce a constraint that all students assigned to a given project must come from the same lab. This constraint ensures that teams have overlapping availability during lab hours, facilitating consistent collaboration. To support this requirement, we introduce a supplementary algorithm that runs prior to Alg.~\ref{algo}, effectively partitioning the set of projects across all the labs before student-to-project assignments are made.

We refer to the lab associated with the students assigned to a project as the project’s \emph{assigned lab}. The supplementary algorithm starts by computing the number of teams needed in each lab by dividing the number of students with an instructor-defined base team size, rounded up. To assign projects to labs, we formulate the problem as a version of the college admissions problem, solvable via the Gale-Shapley algorithm~\cite{gale1962college}, which seeks a stable matching between projects and labs, i.e., a matching without blocking pairs.

The preference list for each lab is constructed by sorting the projects in descending order based on the average project preference score from that lab’s students. Conversely, the preference list for each project is constructed by sorting the labs in descending order of the average \emph{lab skill frequency} of each of the project’s required skills. A skill’s \emph{lab skill frequency} is defined as the percentage of students in that lab who self-report having that skill. Once preference lists are constructed, we match the labs to the projects using the Gale-Shapley algorithm.

While the Gale-Shapley algorithm ensures stability in lab-to-project assignments, it does not account for skill diversity within student teams. As such, once projects are assigned to labs using this method, team formation proceeds using our main algorithm (Alg.~\ref{algo}) within each lab, where skill-based considerations are fully integrated.

\section{Results and Discussion}
\label{sec:results}
To evaluate the effectiveness of our approach, we utilize real-world data gathered from the computer science capstone program at UC Merced. This capstone program is offered every semester as a required culminating experience for senior-level computer science students. In this program, external sponsors submit project ideas, and student teams are subsequently formed to complete these projects over a semester. For team assignments in all semesters, we aimed to maintain reasonable team sizes of three to five students, guided by existing research on optimal team sizes~\cite{Linda-00,Sutherland-14,Susan-09} and the typical scope of capstone projects at UC Merced.

The data analyzed was collected from the Fall 2023 and Spring 2024 semesters. In Fall 2023, the dataset included 16 projects and 68 students, with each project-team pairing being unique, i.e., no project was assigned to more than one student team. In Spring 2024, due to increased enrollment and insufficient availability of unique projects, multiple student teams were assigned to some of the projects. Specifically, Spring 2024 had 22 unique projects, with 5 assigned to two teams each. Thus, the Spring 2024 dataset consisted of 27 projects involving 122 students.

\subsection{Quantitative Findings}
\label{sec:results_findings}
We evaluate our team formation algorithm using two metrics: \emph{Percentage of Skills Fulfilled}, which measures the proportion of required project skills covered by assigned students, and \emph{Average Preference Toward Assigned Project}, which captures the average student preference for their assigned project. Let $K_s \subseteq K$ be the set of skills possessed by student $s$ (for which they assigned a score >1 in the skill survey). Also, let $S_p \subseteq S$ be the set of students assigned to project $p$. The two metrics can be mathematically defined as follows:

\begin{align*}
\text{Percentage of Skills Fulfilled} & = \frac{\left|\left(\bigcup\limits_{s\in S_p} K_s \right)\cap K_p\right|}{|K_p|}\ \\
\text{Average Preference Toward Assigned Project} & = \frac{1}{|S_p|}\sum_{s\in S_p} PS_s(p)
\end{align*}

To evaluate the effectiveness of our approach, we compare its performance across three scenarios. In the first scenario, students are assigned to projects using our algorithm, which balances skill coverage and preference alignment through a dynamic match scoring system described in Sec~\ref{sec:match}. In the second scenario, the instructor manually forms teams using the same survey data, attempting to optimize both skill coverage and student preferences. The third scenario serves as a baseline, in which students are assigned to projects at random without considering any input data. 

\begin{figure}
\centering
\begin{tabular}{cc}
  \begin{minipage}{0.4\linewidth}
    \centering
    \includegraphics[width=\linewidth]{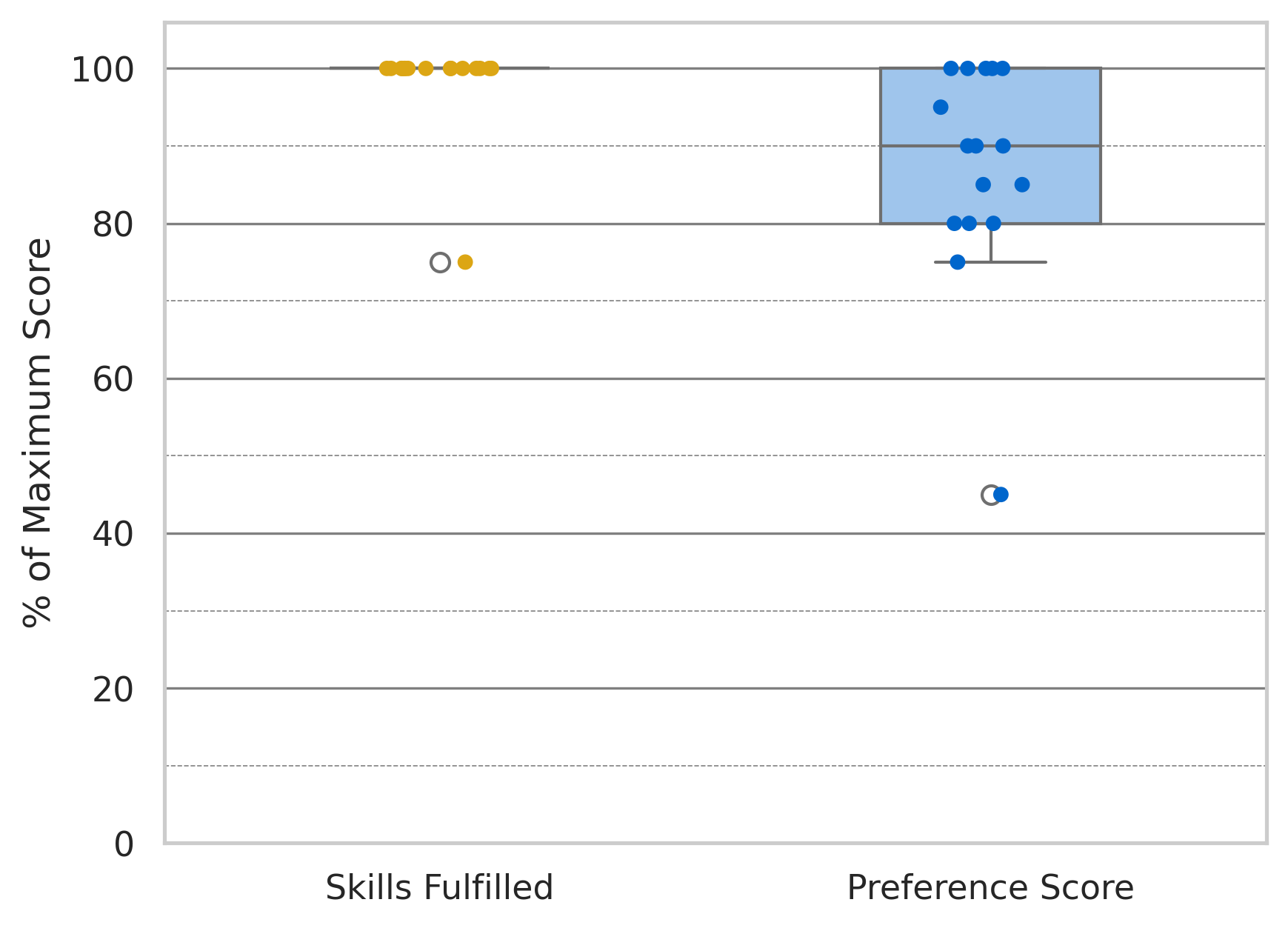}
  \end{minipage} &
  \begin{minipage}{0.4\linewidth}
    \centering
    \includegraphics[width=\linewidth]{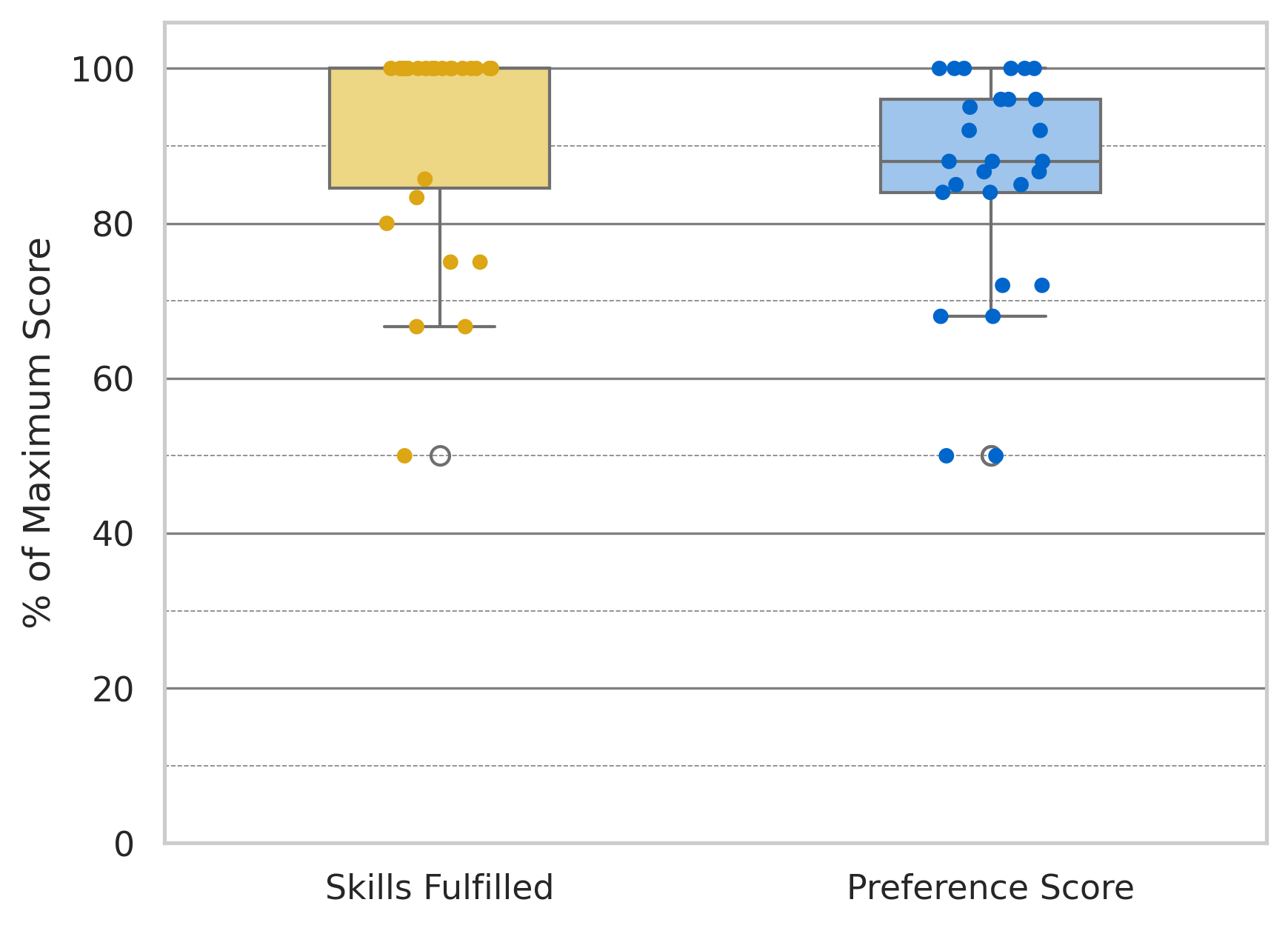}
  \end{minipage} \\ [5pt]
    \multicolumn{2}{c}{
    \begin{minipage}{0.95\linewidth}
      \centering
      \caption*{\scriptsize (a) Ours (Fall 2023 and Spring 2024)}
    \end{minipage}
  } \\
  
  \begin{minipage}{0.4\linewidth}
    \centering
    \includegraphics[width=\linewidth]{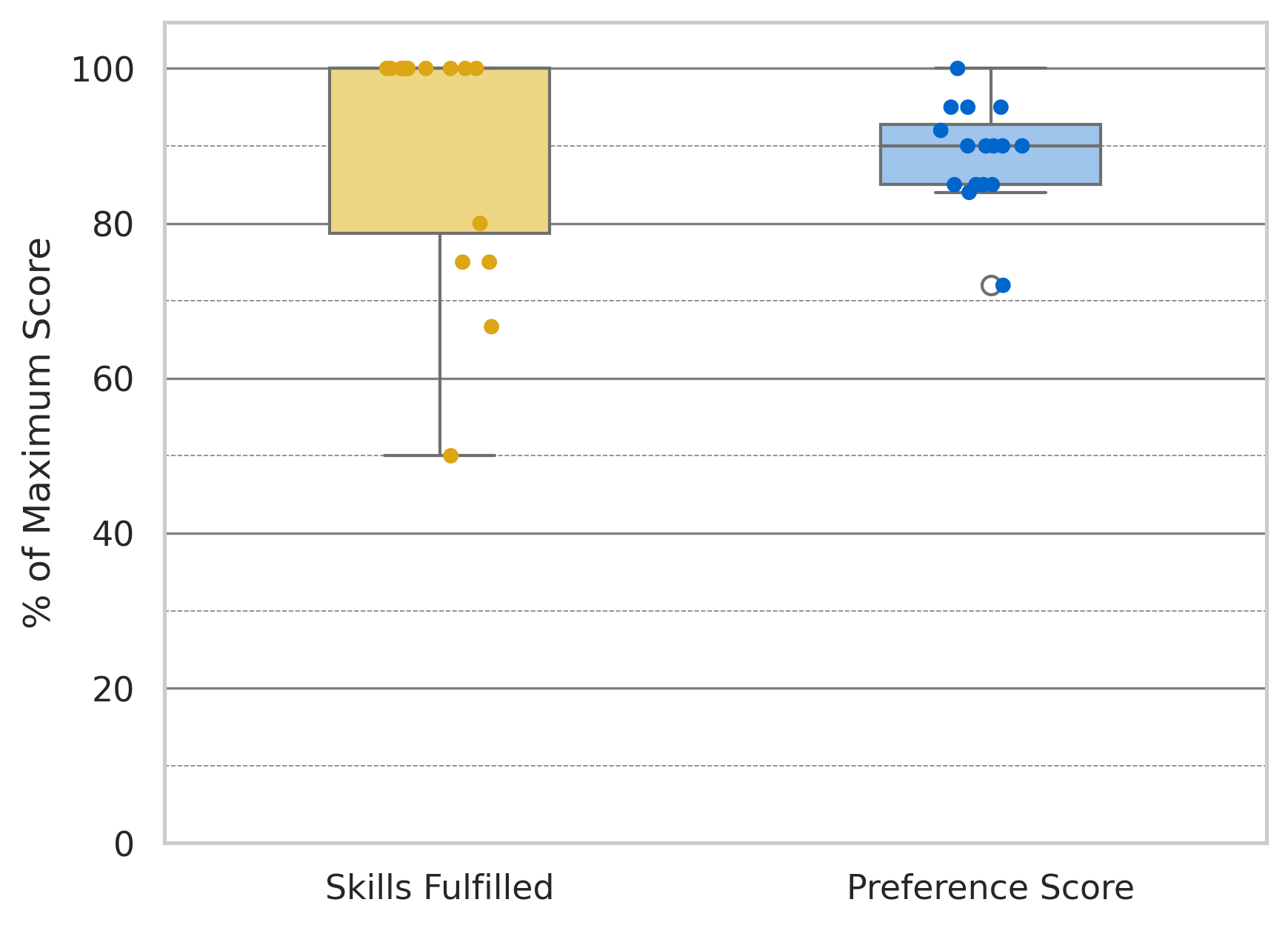}
  \end{minipage} &
  \begin{minipage}{0.4\linewidth}
    \centering
    \includegraphics[width=\linewidth]{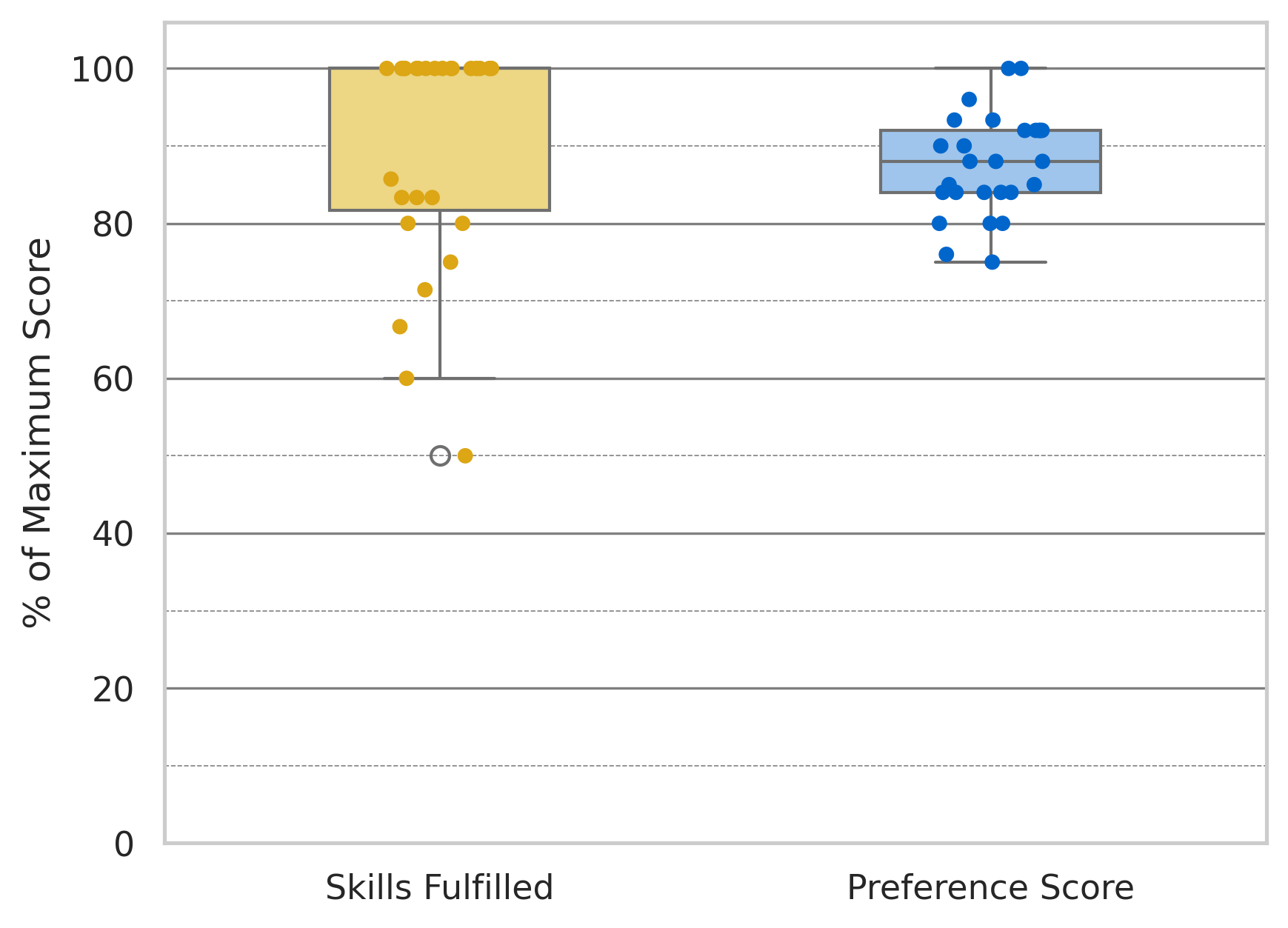}
  \end{minipage} \\ [5pt]
    \multicolumn{2}{c}{
    \begin{minipage}{0.95\linewidth}
      \centering
      \caption*{\scriptsize (b) Manual Assignment (Fall 2023 and Spring 2024)}
    \end{minipage}
  } \\
  
  \begin{minipage}{0.4\linewidth}
    \centering
    \includegraphics[width=\linewidth]{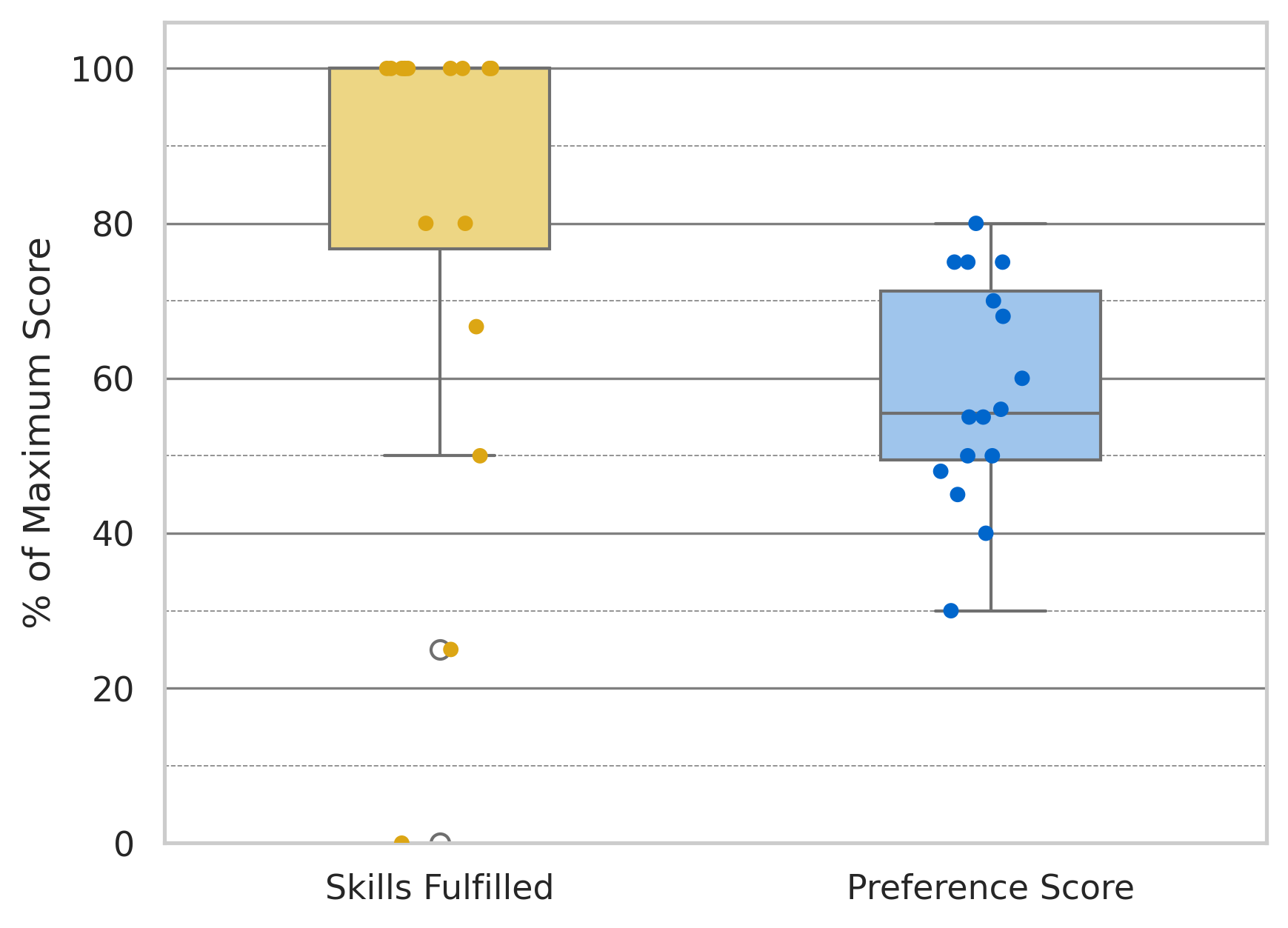}
  \end{minipage} &
  \begin{minipage}{0.4\linewidth}
    \centering
    \includegraphics[width=\linewidth]{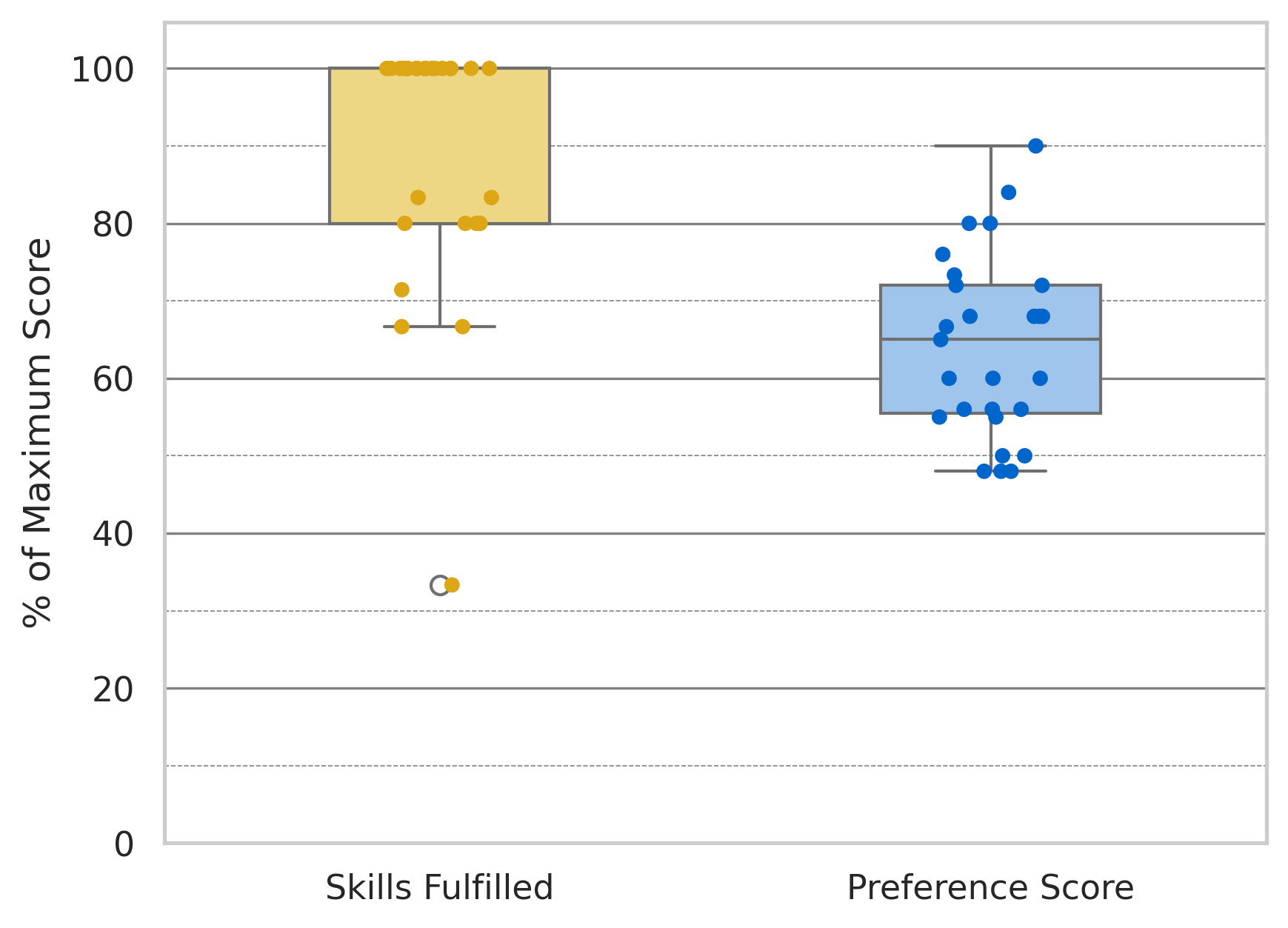}
  \end{minipage} \\ [5pt]
  \multicolumn{2}{c}{
    \begin{minipage}{0.95\linewidth}
      \centering
      \caption*{\scriptsize (c) Random Assignment (Fall 2023 and Spring 2024)}
    \end{minipage}
  } \\

\end{tabular} \\[-10pt]
\captionsetup{labelformat=empty} 
\caption{\scriptsize Fig. 2: Comparison of three team formation methods -- (a) our algorithm, (b) manual, and (c) random -- applied to Fall 2023 (left) and Spring 2024 (right). The ``Skills Fulfilled'' and ``Preference Score" metrics in the plots correspond to the \textit{Percentage of Skills Fulfilled} and \textit{Average Preference Toward Assigned Project} described in text.}
\label{fig:assignment_results}

\end{figure} 

\begin{table}[!ht]
    \begin{minipage}[t]{0.45\textwidth}
        \centering
        \begin{tabular}{|c|c|c|}
            \hline
            (\%) & Fall 2023 & Spring 2024  \\
            \hline
            Ours & \textbf{98.4}  & \textbf{91.9}  \\
            \hline
            Manual & $90.4$  & $89.6$  \\
            \hline
            Random & $81.4$  & $89.8$  \\
            \hline
        \end{tabular}
        \vspace{4mm}
        \captionsetup{labelformat=empty}
        \caption{\scriptsize (A) Percentage of Skills Fulfilled}
        \label{tab:skills_fulfilled}
    \end{minipage}\hspace{17pt}
    \begin{minipage}[t]{0.45\textwidth}
        \centering
        \begin{tabular}{|c|c|c|}
            \hline
            (\%) & Fall 2023 & Spring 2024  \\
            \hline
            Ours & 87.2 & 86.3 \\
            \hline
            Manual & \textbf{88.7}  & \textbf{87.6}  \\
            \hline
            Random & 58.2  & 64.0  \\
            \hline
        \end{tabular}
        \vspace{4mm}
        \captionsetup{labelformat=empty} 
        \caption{\scriptsize (B) Average Preference Toward Assigned Project}
        \label{tab:metrics}
    \end{minipage}
    \captionsetup{labelformat=empty}
    \caption{\scriptsize Tables~\ref{tab:metrics}A and~\ref{tab:metrics}B: Performance metrics of various assignment methods. The values are the average performance across each project. The highest scores for each method are in bold.}
\end{table}

Analysis of our results from Fig~\ref{fig:assignment_results} is summarized in Table~\ref{tab:metrics}. Table~\ref{tab:metrics}A presents the average \emph{Percentage of Skills Fulfilled} for the Fall 2023 and Spring 2024 semesters. Here, our algorithm outperforms both the manual and random assignments across both semesters. Although the manual approach performed better than random assignment, it lagged behind the automated method in effectively meeting project skill requirements. Table~\ref{tab:metrics}B shows the average values of the \emph{Average Preference Toward Assigned Project} computed across all projects for both semesters under consideration. Among the three assignment strategies, the manual approach achieved the highest average \emph{Student Preference Toward Assigned Project} metric, with our algorithm performing competitively. Both the manual and algorithmic approaches substantially outperformed the random baseline.

These findings highlight a key trade-off inherent in manual assignment strategies. Assigning students to projects based on preferences is relatively straightforward, since preference scores are singular, easily comparable values. However, manually optimizing for skill coverage is far more complex. Instructors need to deal with wide variability in both student skill profiles and project requirements, requiring detailed inspection of dozens of individual ratings. This effort imposes a significant cognitive load on the instructor, thereby increasing the likelihood of oversights. On the other hand, our algorithm is specifically designed to address this challenge by simultaneously accounting for both preference and skill alignment through a dynamic, data-driven scoring mechanism. The results reflect the strength of our approach as it surpasses manual assignment in skill coverage while maintaining a competitive preference alignment. Finally, it is worth noting the practical cost of manual assignment. Creating teams manually required over 20 hours of instructor effort to carefully review data and assign students. In contrast, the algorithm produces assignments of higher quality in a matter of milliseconds, demonstrating not only better balance but also high efficiency.

\subsection{Qualitative Findings}
\label{sec:results_qualitative}
We also collected a brief satisfaction survey from 22 previous Capstone students to understand how team formation experiences mapped to perceived fairness, learning, and engagement. This work was approved by our institution’s Institutional Review
Board (IRB). The survey was comprised of the following 7 questions:
\begin{description}
    \item[Q1:] Did the process of assigning teams/projects feel fair?
    \item[Q2:] Overall, how happy are you with the project you got?
    \item[Q3:] Did the team collectively have the needed skills?
    \item[Q4:] Did you personally get to use or grow your skills in the work?
    \item[Q5:] Was the project aligned with what you like?
    \item[Q6:] Did you actually learn new tech or knowledge?
    \item[Q7:] What matters more to you in assigning teams?
\end{description}

\begin{figure}[!ptbh]
    \centering
    \begin{minipage}{0.48\linewidth}
        \centering
        \fboxsep=2pt\fboxrule=0.4pt
        \fbox{\includegraphics[width=0.95\linewidth]{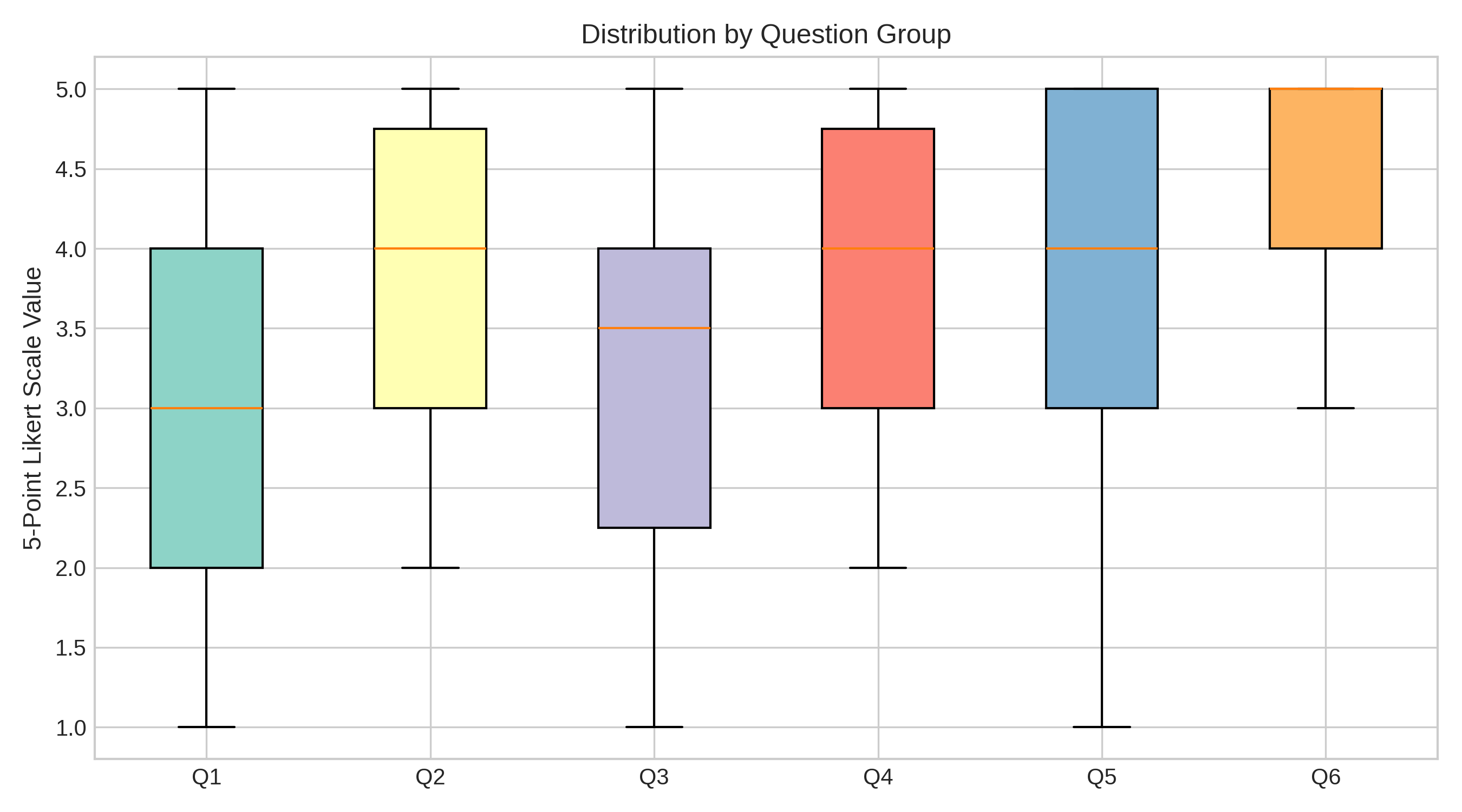}}
        \caption*{\scriptsize (a) Aggregated boxplots for Q1 (fair assignment), Q2 (project satisfaction), Q3 (team skill mix), Q4 (my skills used), Q5 (interest kept me engaged), and Q6 (learned new skills).}
    \end{minipage}\hfill
    \begin{minipage}{0.48\linewidth}
        \centering
        \fboxsep=2pt\fboxrule=0.4pt
        \fbox{\includegraphics[width=0.95\linewidth]{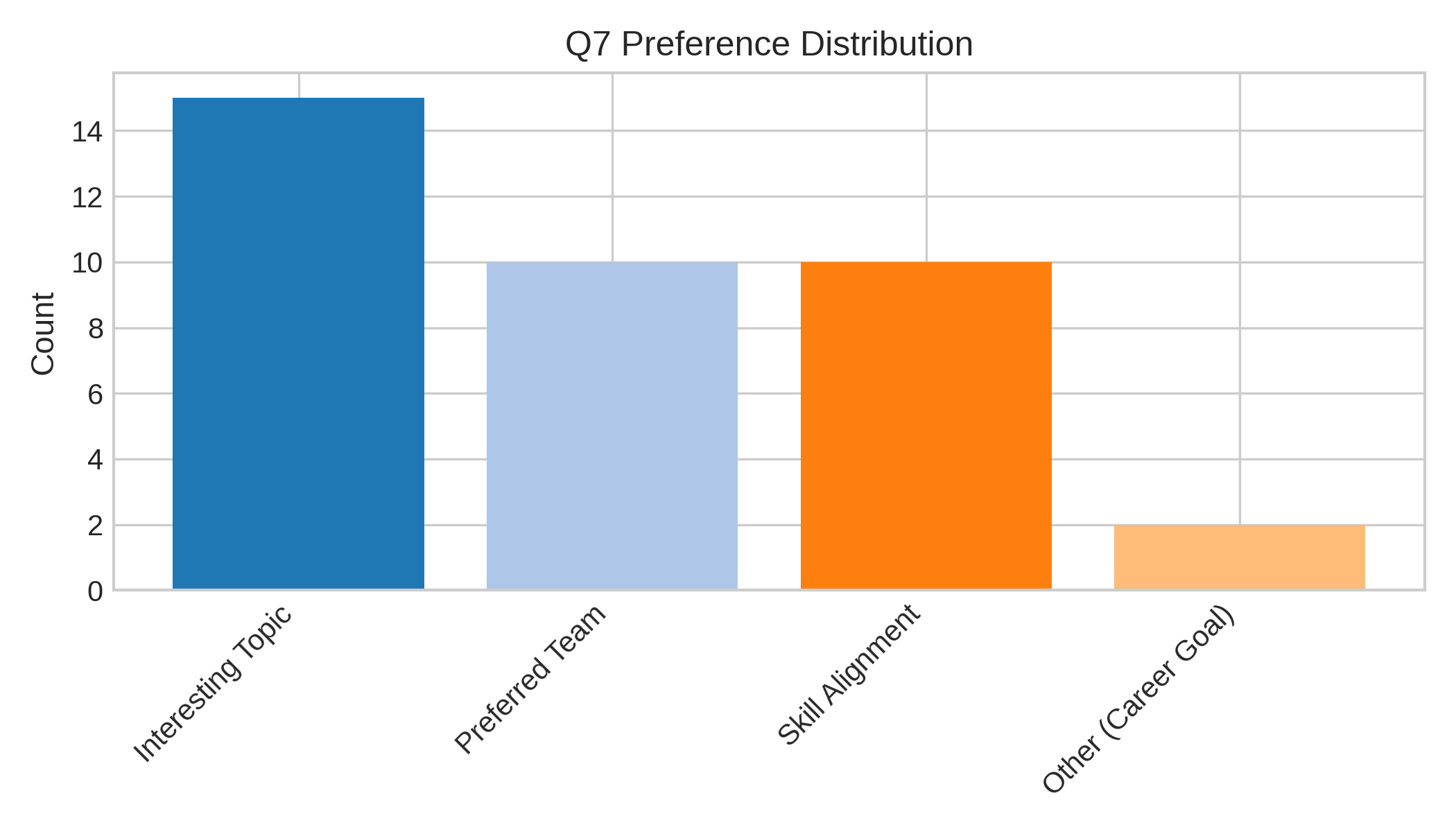}}
        \caption*{\scriptsize (b) Q7 multi-select priorities (counts): Interesting Topic, Skill Alignment, Preferred Team, Other (Career Goal).}
    \end{minipage}
    \\[4pt]
    \captionsetup{labelformat=empty}
    \caption{\scriptsize Fig. 3: Qualitative survey results summarizing distributions across core satisfaction/learning items (a) and student priorities for team assignment (b).}
    \label{fig:qual_plots}
\end{figure}

Fig.~\ref{fig:qual_plots}a shows aggregated boxplots for Q1, Q2, Q3, Q4, Q5, and Q6 (Likert 1--5) and
Fig.~\ref{fig:qual_plots}b visualizes multi-select priorities from Q7. 

Three themes dominated: desire for an interesting topic, preference to work with specific teammates, and aligning work with one’s skills or career direction. Qualitatively, students prioritizing an “interesting topic” sometimes reported lower satisfaction when reality did not match expectations, while those emphasizing “preferred teammates” were often more satisfied but also more likely to note uneven contributions. Together with the boxplots, these findings suggest that managing expectations about project scope, supporting light-weight accountability for contributions, and capturing skill/career signals early can improve both perceived fairness and overall satisfaction.

\section{Conclusions and Future Work}
In this paper we presented a dynamic, data-driven algorithm for forming student teams in capstone and senior design courses that integrates student preferences with project skill requirements. By combining LLM-based skill extraction with dynamic weighting of skills and preferences, our approach addresses limitations of existing tools such as CATME. Evaluation on real course data showed that it improves skill coverage while maintaining high levels of student satisfaction, outperforming random assignment and reducing instructor workload compared to manual assignment.

Despite these advantages several limitations remain. Reliance on self-reported skills and preferences leaves the algorithm vulnerable to misrepresentation. The preference weight parameter $\alpha$ requires manual tuning, and the current method can favor projects with many required skills at the expense of those with rare but critical skills. All skills are currently treated equally, with fulfillment based on a single student’s intermediate-level rating, and the LLM-generated project skills still require manual verification.

Our ongoing work is focused on addressing these issues through adaptive preference weighting, normalization strategies to handle skill rarity, refined fulfillment thresholds, and faculty-validated datasets for improving skill extraction. We are also investigating batch and multi-pass assignment strategies and stable matching variants such as the classified stable matching problem \cite{huang2010classified}. These refinements aim to enhance fairness, scalability, and educational impact, enabling instructors to form teams that are both technically capable and well-aligned with student interests.

\bibliographystyle{splncs04}
\bibliography{references}

\end{document}